\documentclass[aps,pra,showpacs,preprintnumbers,amssymb,amsfonts,floatfix,a4paper,twoside,twocolumn]{revtex4}
\usepackage{amsmath,amssymb,epsfig}
\usepackage{color}

\usepackage{enumerate}

\newcommand{\ket}[1]{ \left|#1\right>}
\newcommand{\bra}[1]{ \left<#1\right|}

\newcommand{\outerp}[2]{ \left|#1\right>\left<#2\right|}	
\newcommand{\expect}[1]{ \left<#1\right>}

\usepackage{soul}
\usepackage{graphicx}
\begin{document}

\title{{\Large{\bf Optical bistability enabled control of resonant light transmission for an atom-cavity system}}}

\author{Rahul Sawant}
\email{rahuls@rri.res.in}
\author{S. A. Rangwala}
\affiliation{%
Light and Matter Physics Group, Raman Research Institute, Sadashivanagar, Bangalore 560080, India}

\begin{abstract}

The control of light transmission through a Fabry-P\'erot cavity containing atoms is theoretically investigated, when the cavity mode beam and an intersecting control beam are both close to specific atomic resonances. A four-level atomic system is considered and its interaction with the cavity mode is studied by solving for the time dependent cavity field and atomic state populations. The conditions for optical bistability of the atom-cavity system are obtained in steady state limit. For an ensemble of atoms in the cavity mode, the response of the intra-cavity light intensity to the intersecting resonant beam is understood for stationary atoms (closed system) and non-static atoms (open system). The open system is modelled by adjusting the atomic state populations to represent the exchange of atoms in the cavity mode, with the thermal environment. The solutions to the model are used to qualitatively explain the observed steady state and transient behaviour of the light in the cavity mode, in Sharma et. al.~\cite{arijit}. The control behaviour with three- and two-level atomic systems is also studied, and the rich physics arising out of these systems, for closed and open atomic systems is discussed.

\end{abstract}
\pacs{37.30.+i, 42.50.Gy, 42.65.Pc.}
\maketitle

\section{Introduction}

A resonantly coupled atom-cavity system can be utilized to explore various possibilities, ranging from linear and nonlinear physics~\cite{arijit,gibbs,WangPRA65,WangPRA65051802,MlynekPRA,LowenauPRL,OrozcoPRA,revLugiatoGibbs,macovei}, single photon-atom cavity systems~\cite{raimond,Haroche,rempe,walther,kimble} all the way to multiple atoms~\cite{Cha03,Her07,Bie10,Wan12,tridib} in a cavity supported mode. 
In this article we theoretically explore the manipulation of the cavity mode (\textit{probe}) light for an atom-cavity system, where the light is on resonance or near resonant with a dipole transition of the atoms contained within the mode volume. The transmission of light through the cavity, on resonance with the atoms, is altered in the presence of another transverse (\textit{control}) beam of light on atomic resonance, which affects the state population of the atoms in the cavity mode. A significant feature of this work is the effect on light transmission when the atoms are stationary within the cavity mode, that is a \textit{closed system} for the atoms and also for an \textit{open system} of atoms, where the cavity mode atoms are exchanged with a reservoir of atoms in mixed ground states. The two cases lead to different experimental behaviour.

The atom-cavity system is modelled and optical bistability of the resonant atom-cavity system is derived. 
The model solves for steady state and transient evolution of this system.
A four-level atomic system is considered initially, and three- and two- level systems are discussed subsequently. The connection to experiments is made by considering the various atomic subsystems for the $^{87}$Rb atom, and the present model provides the requisite insights for a qualitative understanding of the experimental results in Sharma et. al.~\cite{arijit}. The application of this model to laser cooled atoms~\cite{tridib} or ions coupled to a cavity, is discussed. Further, the flow of atoms in and out of the cavity mode is included by modelling the residence time of an atom in the cavity mode, so that the transient behaviour of the atom-cavity systems can be captured. Comparisons with the experiments of Sharma et. al.~\cite{arijit} are made where applicable, and the fact that the model furnishes explanations for the experimentally observed features validates the theoretical model. The model itself is more general that the experimental results~\cite{arijit} and sets the platform for a variety of experiments.   

\section{The atom-cavity system}

\begin{figure}[t]
\centering
\includegraphics[width=8.5cm]{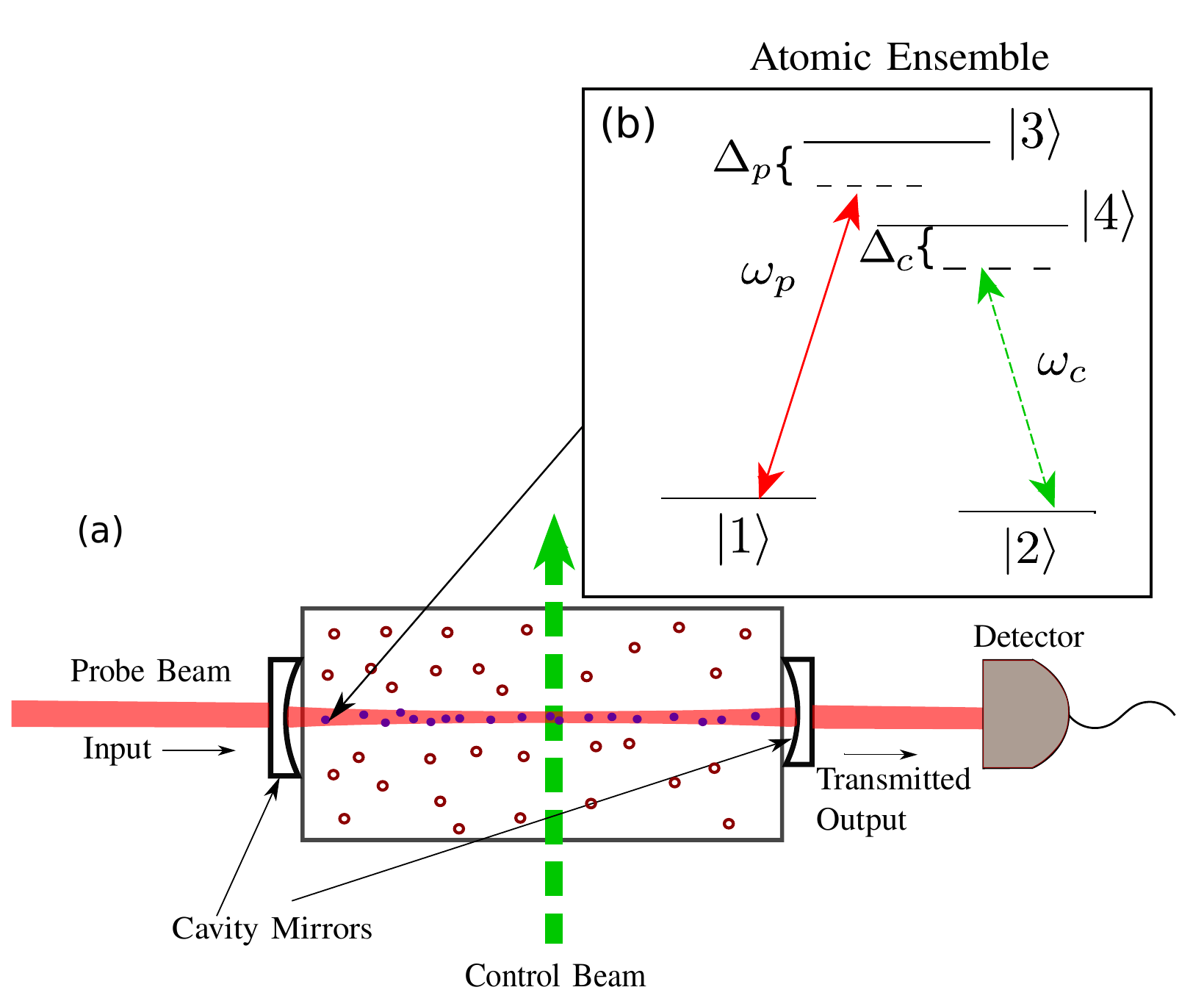}
\caption{(a) Schematic diagram of the system. The atoms are enclosed in a cell, located within the Fabry-P\'erot cavity. The filled circles represent atoms overlapped with the mode and the empty circles represent the background atom vapor. The incident probe beam, on cavity resonance is transmitted through. A transverse control beam (dashed) intersects the cavity mode volume. On atomic resonance, cavity mode light intensity is affected by the presence of the control light beam, which affects the light transmitted through the cavity. The inset (b) shows the atomic level scheme. The probe laser beam is near resonant with transition $\ket{1}\leftrightarrow\ket{3}$ with frequency $\omega_p$ and detuning $\Delta_p$, and the control laser beam is near resonant with $\ket{2}\leftrightarrow\ket{4}$ with frequency $\omega_c$ and detuning $\Delta_c$.}
\label{twobeams}
\end{figure}

Consider a cavity with finesse $F$ and mode volume $V$, which encloses a thermal vapour of atoms. In the model here, we assume that the atomic reservoir is significantly larger than the cavity mode volume. This allows the study of the atom-cavity system in two cases, (a) the closed system, where the atoms contained in the cavity mode are assumed to be stationary and so the same atoms are always coupled to the cavity and (b) the open system, where the atoms enter and exit the cavity mode volume, with the constraint that the total number of atoms within the cavity mode is constant. Initially, the calculations are setup for stationary atoms and the exchange of atoms is incorporated later. 

The schematic conceptualization of the system is illustrated in Fig.~\ref{twobeams}(a). For specificity we consider a medium made up of atoms having two ground states $\ket{1}$ and $\ket{2}$. States $\ket{1}$ and $\ket{2}$ are non degenerate, and cannot decay by any radiative or other process from one to another. The only process by which the ground state populations of the stationary atoms change is via optical transitions, with the excited states. The two ground states are energetically close to each other when compared to first excited state, such that at room temperature the population is split equally between the two ground states. The excited atomic states are $\ket{3}$ and $\ket{4}$, which are non-degenerate. Dipole allowed transitions are possible between both the excited states and the ground states, and so an excited atom can spontaneously emit a photon and populate either one of these ground states. 
Below we consider mechanisms which transfer the populations from one ground state to another, in the presence of the externally applied light fields, for various atomic level subsystems. While four level atoms in a cavity have been previously studied in a ring cavity~\cite{ming,anton}, in our work we use a Fabry-P\'erot cavity. The details of the cavity parameters used are provided in the text, when the first results are quantified.


\section{Closed system of atoms}

\subsection{Four level atoms}

For the four-level stationary atoms, as shown in Fig.~\ref{twobeams} we have an ensemble of atoms placed within the cavity mode. The cavity mode beam (probe beam)  with frequency $\omega_p$, is tuned to the $\ket{1}\leftrightarrow\ket{3}$ resonance. The control beam  with frequency $\omega_c$, is orthogonal to the cavity axis and is tuned to the $\ket{2}\leftrightarrow\ket{4}$ resonance. Identifying the operator $\hat{\sigma}_{ij}\equiv\ket{i}\bra{j}$, we can write the Hamiltonian for the four level atom with eigenstates $\ket{2}$, $\ket{1}$, $\ket{3}$ and $\ket{4}$ and eigenenergies $\hbar\omega_2=0$, $\hbar\omega_1$, $\hbar\omega_3$ and $\hbar\omega_4$, along with the two fields, consistent with the energy level diagram in Fig.~\ref{twobeams}(b) as~\cite{scully,walls}, 
\begin{eqnarray}
\hat{H}&=&\hbar\left(\hat{\sigma}_{11}\omega_{1}+\hat{\sigma}_{33}\omega_{3}+\hat{\sigma}_{44}\omega_{4}\right) \nonumber \\
&& + \hbar g \left(\hat{a}^{\dagger} \hat{\sigma}_{13} e^{i \omega_p t}+ \hat{a} \hat{\sigma}_{31} e^{-i \omega_p t}\right) \nonumber \\
&& + \hbar \left(\Omega^{*} \hat{\sigma}_{24}e^{i \omega_c t}+ \Omega \hat{\sigma}_{42} e^{-i \omega_c t}\right). 
\end{eqnarray}
In the above equation, $g=-\mu_{13}\sqrt{\omega_p /(2 \hbar \epsilon_0 V)}$ is atom-cavity coupling, where $\mu_{13}$ is the transition dipole matrix element between $\ket{3}$ and $\ket{1}$, $\hat{a}$ and $\hat{a}^{\dagger}$ are photon annihilation and creation operators for the probe field, $\Omega=-\mu_{24} |E_c|/\hbar$ is Rabi frequency for control beam, where $\mu_{24}$ is the transition dipole matrix element between $\ket{4}$ and $\ket{2}$, $|E_c|$ is the magnitude of electric field of the control light, and $^{*}$ denotes complex conjugate.  The intensity of control beam is kept constant.  

As we are not in the single photon regime, it is appropriate to consider the expectation values for the field operators in the cavity mode, where $\alpha = \langle a\rangle$ and $\alpha^* = \langle a^{\dagger}\rangle$ correspond to the coherent field, and $\alpha_{in}$ is the input probe field. In general the evolution equation for the expectation value for an operator $\hat{X}$, for the atomic system under consideration is written as
\begin{equation}
\frac{d\langle\hat{X}\rangle}{dt} = \frac{i}{\hbar} \langle [\hat{H},\hat{X}]\rangle,
\label{hisen}
\end{equation}
where, $[\hat{H},\hat{X}]$ is the commutator of $\hat{X}$ with the Hamiltonian, $\hat{H}$. For the atom-cavity system defined above, with the restrictions imposed on the four-level system, the evolution equations for the atomic states and the cavity field, after including spontaneous emission rates and cavity loss terms phenomenologically, result in the reduced set of coupled differential equations~\cite{scully,walls,albert,albert1},
\begin{eqnarray}
\dot{\alpha} &=& \sqrt{\frac{\kappa_1}{\tau_c}} \alpha^{in}_p - \kappa_t \alpha  - i g N_{at} \rho_{13} \nonumber \\
\dot{\rho}_{13} &=& -\left(\gamma_{13} + i \Delta_p \right) \rho_{13} + i g \alpha (\rho_{33} - \rho_{11})  \nonumber \\
\dot{\rho}_{33} &=&  -\Gamma \rho_{33} + ig (\alpha^{*} \rho_{13} - \alpha \rho_{13}^{*} ) \nonumber \\
\dot{\rho}_{11} &=&  \frac{\Gamma}{2} \ \rho_{44} + \frac{\Gamma}{2} \ \rho_{33} - ig (\alpha^{*} \rho_{13} - \alpha \rho_{13}^{*} )  \nonumber \\
\dot{\rho}_{24} &=& -\left(\gamma_{24} + i \Delta_c \right) \rho_{24} + i \Omega (\rho_{44} - \rho_{22})  \nonumber \\
\dot{\rho}_{44} &=&  -\Gamma \rho_{44} + i  (\Omega^{*} \rho_{24} - \Omega \rho_{24}^{*} )\nonumber \\
\dot{\rho}_{22} &=&  \frac{\Gamma}{2} \ \rho_{44} + \frac{\Gamma}{2} \ \rho_{33} - i (\Omega^{*} \rho_{24} - \Omega \rho_{24}^{*}). 
\label{4leveleq}
\end{eqnarray}
Here, $\alpha$ is the intra-cavity field, $\alpha_p^{in}$ is the probe field, $\kappa_1$ is transmission rate related to transmission coefficient ($T$) of input mirror given by, $\kappa_1 = \frac{T}{2\tau_c}$, where $\tau_c$ is round trip time of the photon in the cavity, and $\kappa_t$ the total loss rate of cavity. In addition, $\Gamma$ is the spontaneous emission decay rate from the excited states, $\gamma_{13}$ and $\gamma_{24}$ are the decoherence rates for coherence between states $\{\ket{3},\ket{1}\}$ and $\{\ket{4},\ket{2}\}$ respectively. The probe laser detuning from the $\ket{1}\leftrightarrow\ket{3}$ transition is $\Delta_p= (\omega_3 - \omega_1) - \omega_p$ and the control laser detuning from the $\ket{2}\leftrightarrow\ket{4}$ transition is $\Delta_c= (\omega_4 - \omega_2) - \omega_c$. The probe is resonant with the cavity ($\omega_p=\omega_{ca}$).

The objective throughout this work is to solve for the intra-cavity light intensity $\alpha$, when the states of the atoms interact with the light fields. In this case the application of the cavity mode laser $\omega_p$ ($\ket{1}\leftrightarrow\ket{3}$) and the transverse control laser $\omega_c$ ($\ket{2}\leftrightarrow\ket{4}$) couples the ground states of the atoms in a manner that shifts the state populations in the atoms. This in turn affects the intensity of the light in the cavity mode. The entire coupled system of Eqns.~\ref{4leveleq} can be solved to obtained $\alpha$, with realistic parameters for the atom and cavity.

For the atom, we have $\rho_{13} = e^{i\omega_pt}\expect{\outerp{1}{3}}$ and $\rho_{24} = e^{i\omega_ct}\expect{\outerp{2}{4}}$ for the coherences, and $\rho_{nn} = \expect{\outerp{n}{n}}$ for diagonal terms representing the state populations. We neglect collisional dephasing for all transitions, and set $\gamma_{24} = \gamma_{13} =\frac{\Gamma}{2}$ in Eqns.~\ref{4leveleq}. $N_{at}$ is total number of atoms coupled to the cavity. In the $\dot{\alpha}$ equation, the third term represents the loss of intracavity light due to the presence of spontaneously emitting resonant atoms in the cavity and is proportional to no. of atoms ($N_{at})$. \\

\begin{figure}[t]
\centering
\includegraphics[width=7.5cm]{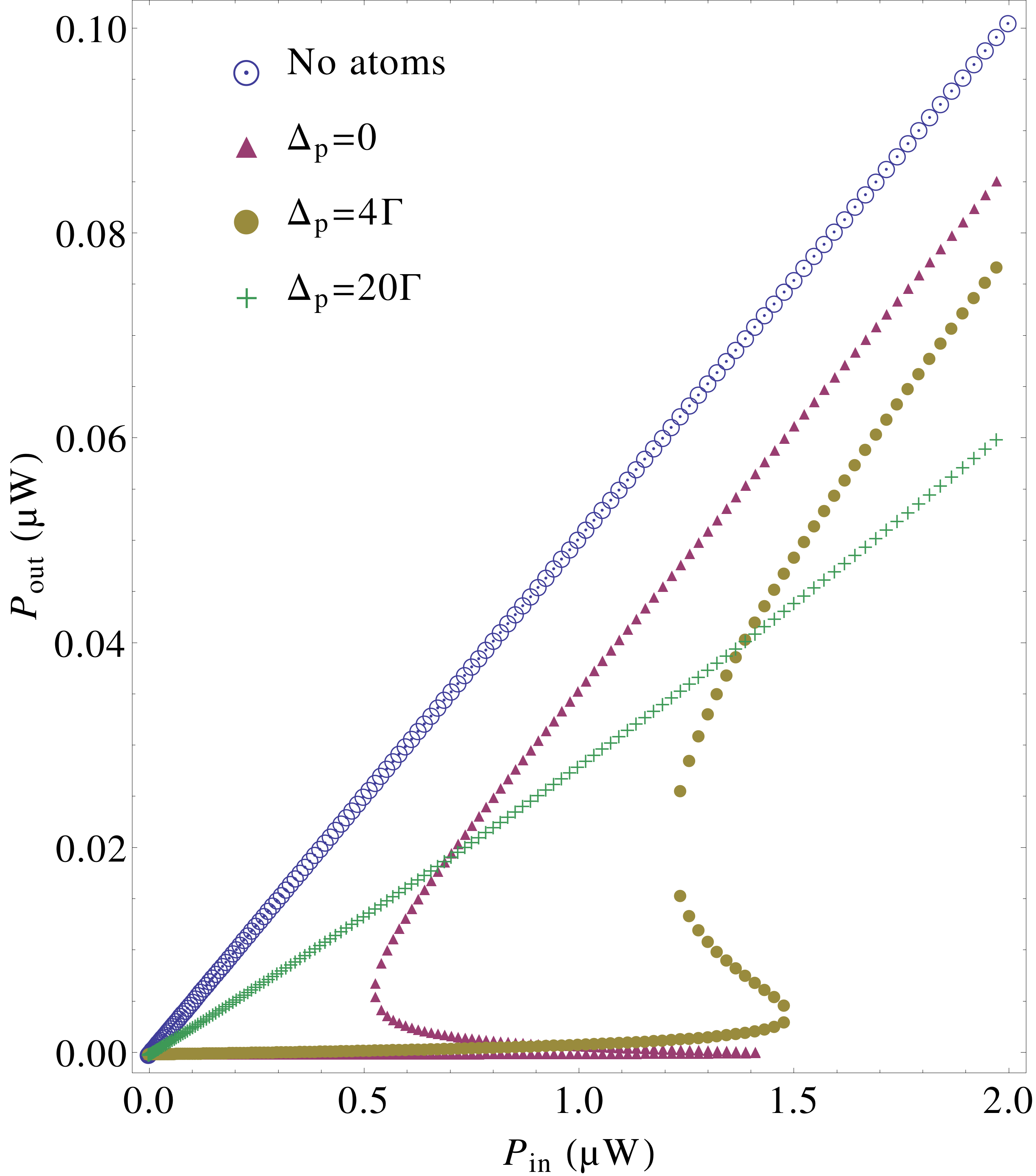}
\caption{Transmitted power of cavity versus input power after solving Eqn.~(\ref{alpha4level}) numerically and converting to SI units. The different curves correspond to different detunings of the probe laser from the atomic resonance. When there are no atoms, then the response is linear. As $\Delta_p$ changes, the nature of the hysteresis and therefore the region for bistable behaviour changes. All curves are for control power corresponding to $\Omega/2\pi = -50$ Ghz and detuning $\Delta_c = 0$}
\label{}
\label{4bistable}
\end{figure}

In steady state, $\dot{\alpha}=0$ and $\dot{\rho}_{mn}=0, \forall(m,n)$ and Eqns.~(\ref{4leveleq}) becomes a set of linear equations which can be solved algebraically. Eliminating the atomic variables we get a nonlinear equation of degree 3 in $\alpha$,
\begin{eqnarray}
\sqrt{\frac{\kappa_1}{\tau_c}} \alpha^{in}_p - \kappa_t \alpha  - \kappa_{at} \alpha = 0,
\label{alpha4level}
\end{eqnarray}
where $\kappa_{at}$ is decay rate of cavity field due to the atoms and is given by,
\begin{eqnarray}
\kappa_{at} =  \frac{2N_{at} g^2  (\Gamma -2 i \Delta_ p) }{|\alpha|^2 g^2\left[(\frac{\Gamma}{|\Omega|})^2+16+4 (\frac{\Delta_c}{|\Omega|})^2\right]+ \left(\Gamma ^2+4 \Delta_p^2\right)}.\nonumber
\end{eqnarray}  
The quadratic dependence on $\alpha$ in the denominator turns Eqn.~(\ref{alpha4level}) into a cubic equation. The negative sign in front of the third term represents the interaction of cavity field with atoms is lossy due to spontaneous emission. The solution of Eqn.~(\ref{alpha4level}) allows the existence of 3 steady state values of $\alpha$ for same $\alpha_{in}$, as shown in Fig.~\ref{4bistable}.

The part of the solution to Eqn.~\ref{alpha4level} with negative slope is an unstable solution~\cite{boyd} of the cubic equation. Experimentally, the system transits from one stable solution to another, therefore exhibiting bistability in the regime where multiple solutions exist for a given value of $\alpha_{in}$. In Fig.~\ref{4bistable}, we have converted the cavity field into power, for comparison with experiments. This is achieved by multiplying intra-cavity intensity $\frac{c\epsilon_0}{2}|\alpha|^2 \left(\frac{\hbar \omega _{ca}}{2V\epsilon_0}\right)$ by transmission coefficient $2\kappa_1\tau_c $ and area of beam $A = 6.2 \times 10^{-2}$ mm$^2$. Here $\sqrt{\frac{\hbar \omega _{ca}}{2V\epsilon_0}}$ is the proportionality constant between electric field operator inside cavity and the operators $\hat{a}$ and $\hat{a}^{\dagger}$, $V$ is volume of cavity mode and $\omega_{ca}$ is cavity resonance frequency. For the calculation the cavity parameters are $\kappa_t =4.38$ MHz, $\kappa_1 = 0.492$ MHz, $\tau_c = 0.305$ ns and $N_{at} = 5 \times 10^4$. The values for atomic transitions are based on the ground states ($\ket{1}$, $\ket{2}$) correspond to $5^2S_{1/2}(F=2,F=1)$ and excited states ($\ket{3}$, $\ket{4}$) correspond to $5^2P_{3/2}(F=1,F=2)$ respectively of $^{87}$Rb D$_2$ transition and $\frac{\Gamma}{2}\pi = 6.06 $ MHz~\cite{steck}. $\Omega/2\pi = -50$ Ghz corresponding to $1 \mu$W power of control beam, $g/2\pi = - 40$ kHz. The parameter values above are fairly representative of our generic experimental laboratory conditions~\cite{tridib}.

\begin{figure}[t]
\centering
\includegraphics[width=7.5cm]{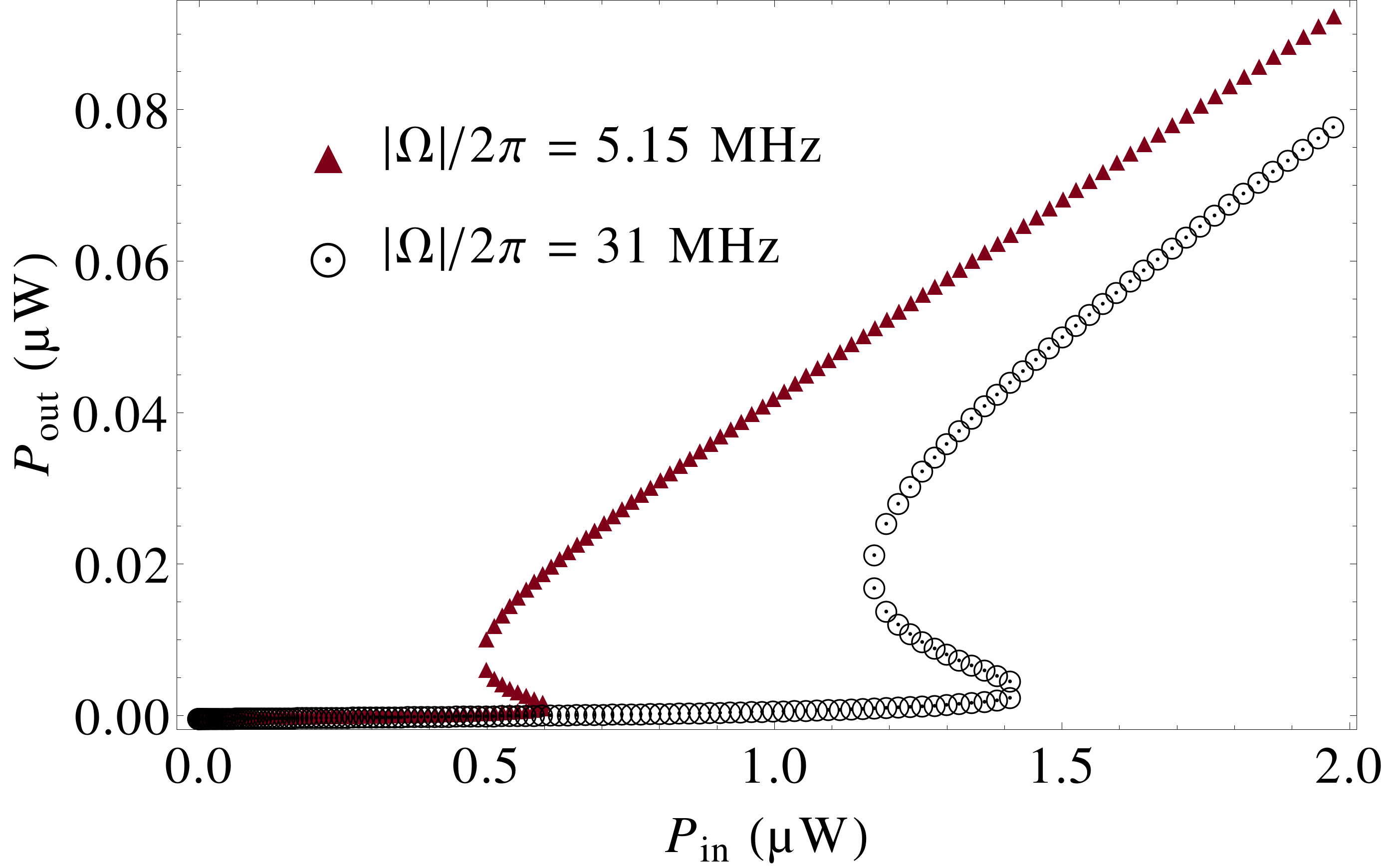}
\caption{Bistable curves for two different control beam powers with $\Delta_p = 2 \Gamma $.}
\label{4shift}
\end{figure}

The cavity output power as a function of input power, in the presence of transverse control light field with $\Omega/2\pi = -50$ Ghz is illustrated in Fig.~\ref{4bistable}.  From the figure, we infer that as the detuning of probe beam from the atomic resonance increases, the region of bistability decreases and beyond a point, for very large detunings, the system reduces to the one where no resonant atoms are present in the cavity and the empty cavity response is obtained. This is clear from Eqn.~(\ref{alpha4level}), in the limit of large $\Delta_p$.

For large control laser power represented by $\Omega/2\pi \approx -50$ Ghz, $16 \gg (\Gamma/\Omega)^2 \ge (\Delta_c/|\Omega|)^2$, the loss in cavity light becomes largely independent of the probe power as can be inferred from above condition and Eqn.~(\ref{alpha4level}). However, for $\Omega$ of the order of $\Gamma$ the loss of cavity light becomes dependent on control power as seen in Fig.~\ref{4shift}. In this range the intra-cavity probe light intensity can be significantly altered by changing the ratio of control power to cavity input power $P_{c}/P_{in} \approx 10^{-8}$. This shows that a very weak transverse intensity of light, under suitable conditions is able to control the intra-cavity light intensity of the probe beam. This regime of control is available only in the region where the bistability is manifested. When the intra-cavity intensity is high, and when $\Omega \gg \Gamma$ we obtain a saturated regime, where adding more power to probe or control beam does not alter the intra cavity field significantly.

\begin{figure}[t]
\centering
\includegraphics[width=7.5cm]{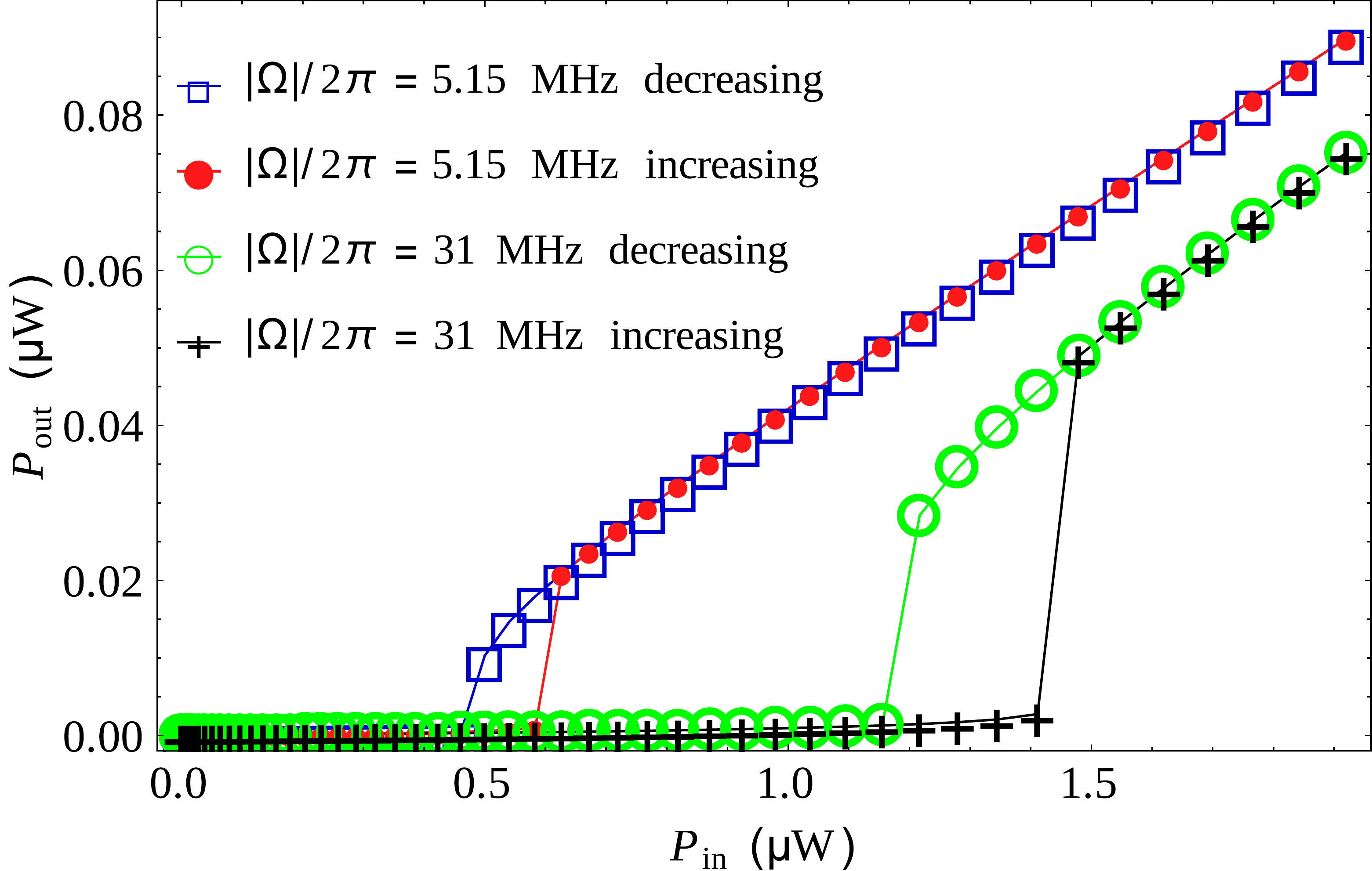}
\caption{Hysteresis for the two different control beam powers in Fig.~\ref{4shift}, with probe beam detuning $\Delta_p = 2 \Gamma $. The hysteresis shown here is a result of following the change in output as the input is increased and decreased adiabatically.}
\label{shiftdyn}
\end{figure}

The set of partial differential equations (Eqns.~(\ref{4leveleq})) can be solved in a Mathematica Notebook numerically, by adiabatically increasing and then decreasing the input power of probe beam while keeping the power of the control beam at a given value. In this case, the change in the output power can be followed by tracking the intra-cavity field $\alpha$ and converting it into a transmitted output intensity using the cavity $\kappa_1$. The resulting intensity of the transmitted light is seen to exhibit a hysteresis as seen in Fig~.\ref{shiftdyn}. Here, as the control power is increased, the hysteresis features moves to higher values of the probe light intensity. The observed hysteresis is consistent with the steady state solutions that exhibit bistability and the transmitted light intensity follows the positive slopes of the bistability solution.


\subsection{Three level atoms} \label{3levelsec}

\begin{figure}[b]
\centering
\includegraphics[width=7.5cm]{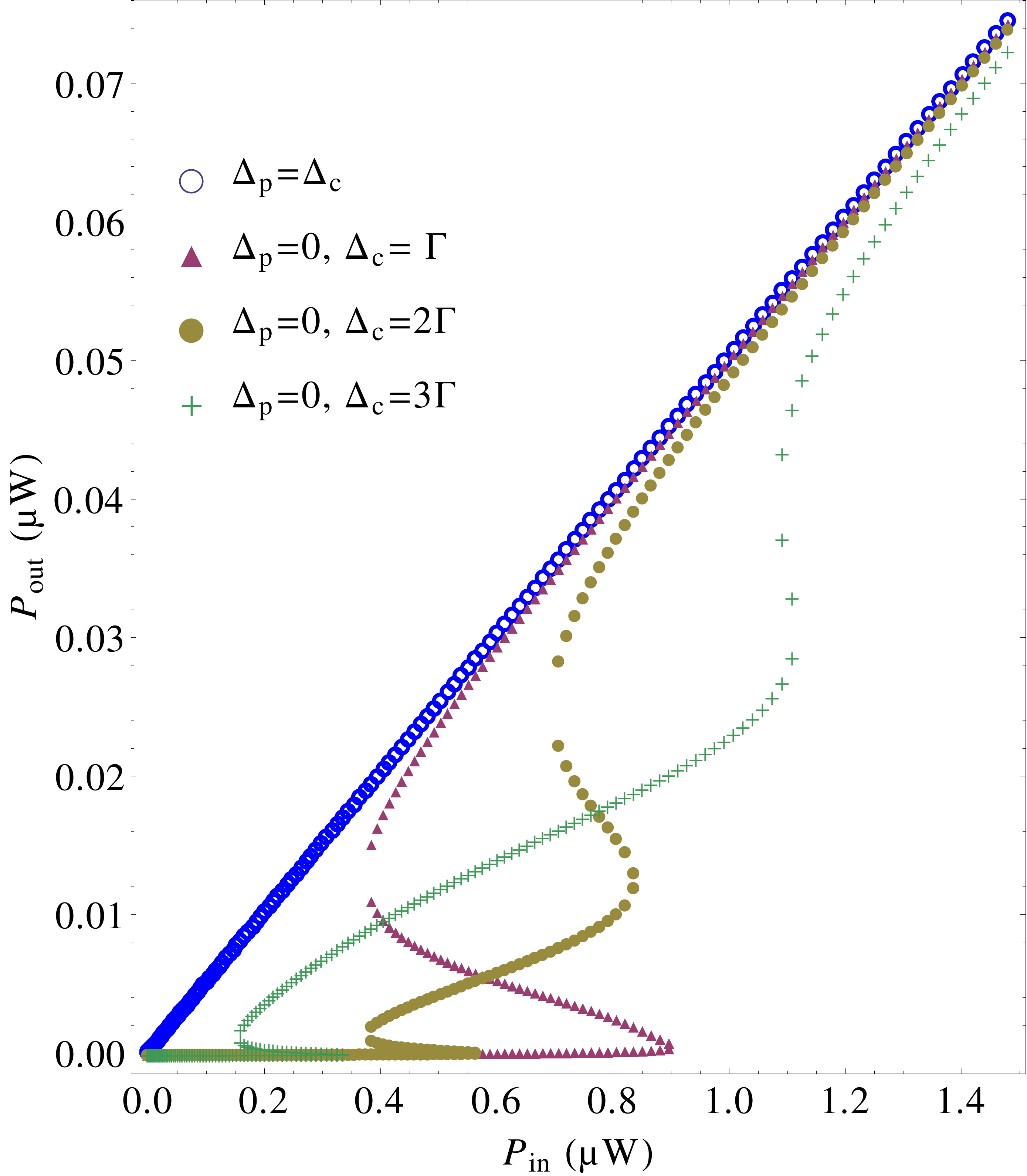}
\caption{Bistability in case of a three level system for different values of detuning. Here $|\Omega|/2\pi = 5.15$ MHz. Unlike the 4 level system the bistable curves have two distinct regions of negative slope in some cases. The four curves illustrate the different cases of probe and control detuning, as specified in legend.}
\label{3levelbi}
\end{figure}

The three-level atom is easily adapted from the earlier four-level atom by eliminating the $\ket{4}$ excited state. The probe beam frequencies and detunings remain the same while the control beam couples $\ket{2}\leftrightarrow\ket{3}$, with detuning $\Delta_c= (\omega_3 - \omega_2) - \omega_c$. The resulting three level system has two ground states and thus is a $\Lambda$ system. The rate equations can be derived in a manner analogous to that of Eqns.~(\ref{4leveleq}), and can be written as,
\begin{eqnarray}
\dot{\alpha} &=& \sqrt{\frac{\kappa_1}{\tau_c}} \alpha^{in}_p - \kappa_t \alpha  - i g N_{at} \rho_{13} \nonumber \\
\dot{\rho}_{13} &=& -\left(\gamma_{13} + i \Delta_p \right)\rho_{13} + i g \alpha (\rho_{33} - \rho_{11}) - i \Omega \rho_{12}  \nonumber \\
\dot{\rho}_{11} &=&  \frac{\Gamma}{2} \ \rho_{44} + \frac{\Gamma}{2} \ \rho_{33} - ig (\alpha^{*} \rho_{13} - \alpha \rho_{13}^{*} )  \nonumber \\
\dot{\rho}_{33} &=&  -\Gamma \rho_{33} + ig (\alpha^{*} \rho_{13} - \alpha \rho_{13}^{*} )+ i  (\Omega^{*} \rho_{23} - \Omega \rho_{23}^{*} ) \nonumber \\
\dot{\rho}_{23} &=& -\left(\gamma_{23} + i \Delta_c \right) \rho_{23} + i \Omega (\rho_{33} - \rho_{22}) -i \alpha g \rho_{12}^* \nonumber \\
\dot{\rho}_{12} &=&  -\left(\gamma'_{12} + i [\Delta_c-\Delta_p] \right) \rho_{12}+i\alpha g\rho_{23}^*-i\Omega^* \rho_{13}\nonumber \\  
\dot{\rho}_{22} &=&  \frac{\Gamma}{2} \rho_{33} - i(\Omega^{*} \rho_{23} - \Omega \rho_{23}^{*} ),
\label{3leveleq}
\end{eqnarray}
where $\gamma_{13}=\gamma_{23}=\frac{\Gamma}{2}$, and $\gamma'_{12}$ is the collisional dephasing, between the ground states. This is the general description when the atoms are in motion, but still coupled to the cavity mode at all times. For the three-level scheme, we consider the complete set of evolution equations, unlike the four-level case. The coherence term between the ground states $\ket{1}$ and $\ket{2}$ plays a role here due to the common excited state $\ket{3}$.  All other notations are the same as those defined alongside Eqns~(\ref{4leveleq}), in the four-level case.
Similar to the four-level case the steady state equation for $\alpha$ can be calculated, yielding an equation in $\alpha$ of the form
\begin{eqnarray}
\sqrt{\frac{\kappa_1}{\tau_c}} \alpha^{in}_p - \kappa_t \alpha  - \tilde{\kappa}_{at} \alpha =0,
\label{alpha3level}
\end{eqnarray}
where $\kappa_1$ and $\kappa_t$ have been defined earlier and $\tilde{\kappa}_{at}$ is decay rate of cavity field from atoms given
by,
\begin{equation}
\tilde{\kappa}_{at}= \frac{T_1\alpha+T_2|\alpha|^2\alpha}{T_3|\alpha|^6+T_4|\alpha|^4+ T_5 |\alpha|^2 +T_6},
\label{3LAtLoss}
\end{equation}
which results in a seventh power equation in $\alpha$, giving rise to interesting consequences. The values of $T_1, ..., T_6$ are given in the appendix.

\begin{figure}[b]
\centering
\includegraphics[width=7.5cm]{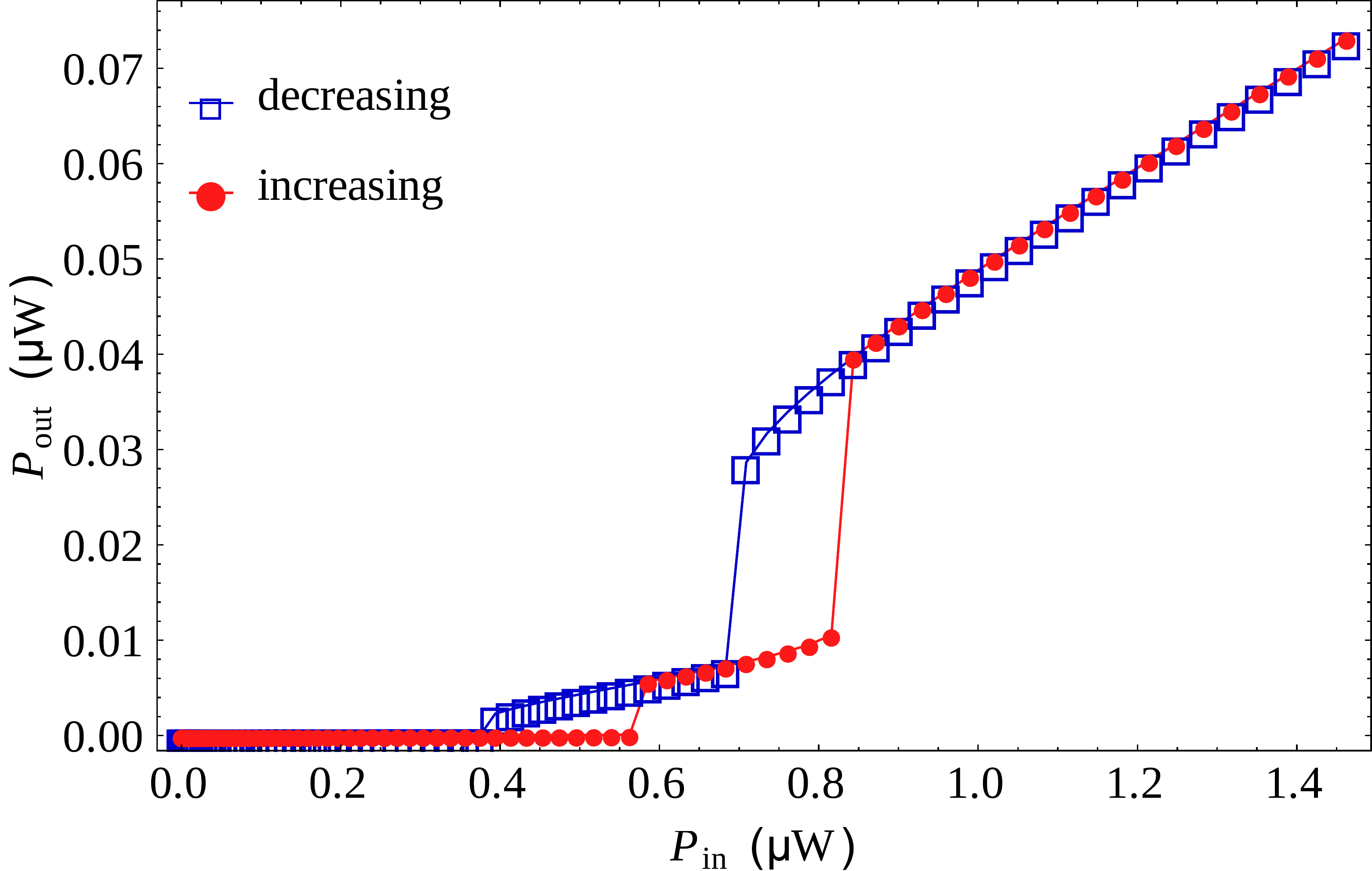}
\caption{Double hysteresis for the case of $\Delta_p=0$, $\Delta_c =2\Gamma$ and $|\Omega|/2\pi = 5.15$ MHz in Fig.~\ref{3levelbi}. The increasing and decreasing legend represents the sense of the change in the input probe power.}
\label{3levelhys}
\end{figure}

For a specific case of $g \ll \Omega,\Gamma$, above solution reduces to the one derived in~\cite{zou-zhu} which does not result in hysteresis. The seventh power of $\alpha$ in Eqn.~(\ref{3LAtLoss}) gives rise to an input-output relation when the two light frequencies are close to resonance, which is far richer than the case of the four level system, within the approximations made. A seventh order cavity field solution can result in multi-stability and multiple bistability as seen in Fig.~\ref{3levelbi} and Fig.~\ref{3levelhys}. 

When the gas of atoms is dilute and stationary, the assumption $\gamma'_{12} \rightarrow 0$ holds and equal detuning i.e $\Delta_p = \Delta_c$ implies that $\tilde{\kappa}_{at} = 0$, resulting in electromagnetically induced transparency (EIT)~\cite{EITrev}. This results in Eqn.~(\ref{alpha3level}) becoming linear in $\alpha$ due to the fact that the medium becomes transparent in steady state and so the transmitted light depends only on cavity loss. Hence there is no bistability, as has been observed experimentally~\cite{joshi}. For $\Delta_p\ne\Delta_c$ we see bistability, as shown in Fig.~\ref{3levelbi}. Depending on the system parameters, the nature of the solution can be altered such that the adiabatic solution of the curve can exhibit multiple bistable behaviour. This is seen in Fig.~\ref{3levelbi} and Fig.~\ref{3levelhys}.

\section{Open system of atoms}

\subsection{Four level atoms} \label{4levelopensec}

So far the atoms that have been considered are static, i.e. they do not flow in and out of the cavity mode. We now consider a situation where the resonant atoms are swapped in and out of the cavity mode, while maintaining the number of atoms within the mode constant with time. Such a system describes the case when the atomic reservoir is much larger than the volume of the cavity mode, and represents an open system of atoms coupled to the cavity. Experimentally this is satisfied for the schematic illustrated in Fig.~\ref{twobeams}. For such a system, the steady state behaviour of the atom-cavity system remains the same while the transient response of the four level atom-cavity system is significantly affected. Below we formulate the atom-cavity problem by creating a simple exchange	 model where atoms are exchanged between the cavity mode and the reservoir. In this case the switching response of light through the cavity can be studied and connected to the experiments in Sharma et. al \cite{arijit}.

The principle difference with respect to the stationary atom model is that the state of the ensemble (density matrix) in the cavity mode changes, as the atoms interacting with the cavity go out of the cavity mode and background gas atoms are added to the ensemble of atoms inside the cavity. Consider the case when the flow rate of atoms from the cavity mode is $R_f$. After time $t$ the fraction of atoms remaining are $e^{-R_f t}$. When $N_1$ atoms with state $\rho_1$ are mixed with, $N_2$ atoms with state $\rho_2$, the statistical mixture of both gives density matrix,
$$\rho=\frac{1}{N_1 + N_2}(N_1\rho_1+N_2\rho_2).$$ 
If $\rho(t)$ is the density matrix at time $t$ the density matrix after small time $\tau$ is 
$$\rho(t+\tau) = e^{-R_f \tau}\rho(t) + (1-e^{-R_f \tau})\rho^{0},$$
where $\rho^{0}$ is density matrix of external atoms entering the cavity mode. For small $\tau \ll \{1/R_f,\tau_c , 1/\Gamma\}$, $e^{-R_f\tau} \approx 1-R_f \tau$ and $(1-e^{-R_f \tau})\approx R_f\tau$. Therefore,
$$\frac{\rho(t+\tau) - \rho(t)}{\tau} = - R_f\rho(t) + R_f \rho^{0}$$ 
and in the limit $\tau \rightarrow 0$, 
$$\frac{d \rho(t)}{dt} = R_f [\rho^{0}-\rho(t)].$$
For four-level atoms the equations are same as eqns.~\ref{4leveleq} with addition of term $R_f(\rho^{0}_{nm}-\rho_{nm})$ on right hand side for atomic operators $\rho_{nm}, \forall{(n,m)}$. The corresponding set of equations with the exchange incorporated is reproduced in the Appendix. 

For an atom at room temperature $\rho^0_{44}=\rho^0_{33}=\rho^0_{13}=\rho^0_{24} \approx 0 $ and $\rho^0_{22}=\rho^0_{11}\approx 0.5$. 
In such a system of atoms the steady state loss of cavity field due to interaction with the atoms, represented by $\kappa_{at}$ in the intracavity field eqn~(\ref{alpha4level}), when control light is off (i.e. $\Omega=0$) is,
\begin{eqnarray}
\kappa_{at}=\frac{N_{at}g^2 R_f A (B -2 i \Delta_p)}{2 |\alpha|^2 g^2 B C+R_f A \left(4 \Delta_p^2+ B^2\right)},
\label{4levelflow}
\end{eqnarray} 
where we identify $A=\Gamma +R_f$, $B=\Gamma +2 R_f$ and $C=\Gamma + 4 R_f$. Thus the system exhibits bistable behaviour, both with and without the presence of the control beam. Here the saturated sample is continuously replaced by a thermal sample hence there is continuous supply of fresh atoms in ground state $\ket{1}$, which is equivalent to the action of the control beam for the stationary atom case. When the control beam is incident on a small fraction of atoms we see a shift in bistability, as shown in Fig.~\ref{4levelflow}.  
\begin{figure}[t]
\centering
\includegraphics[width=7.5cm]{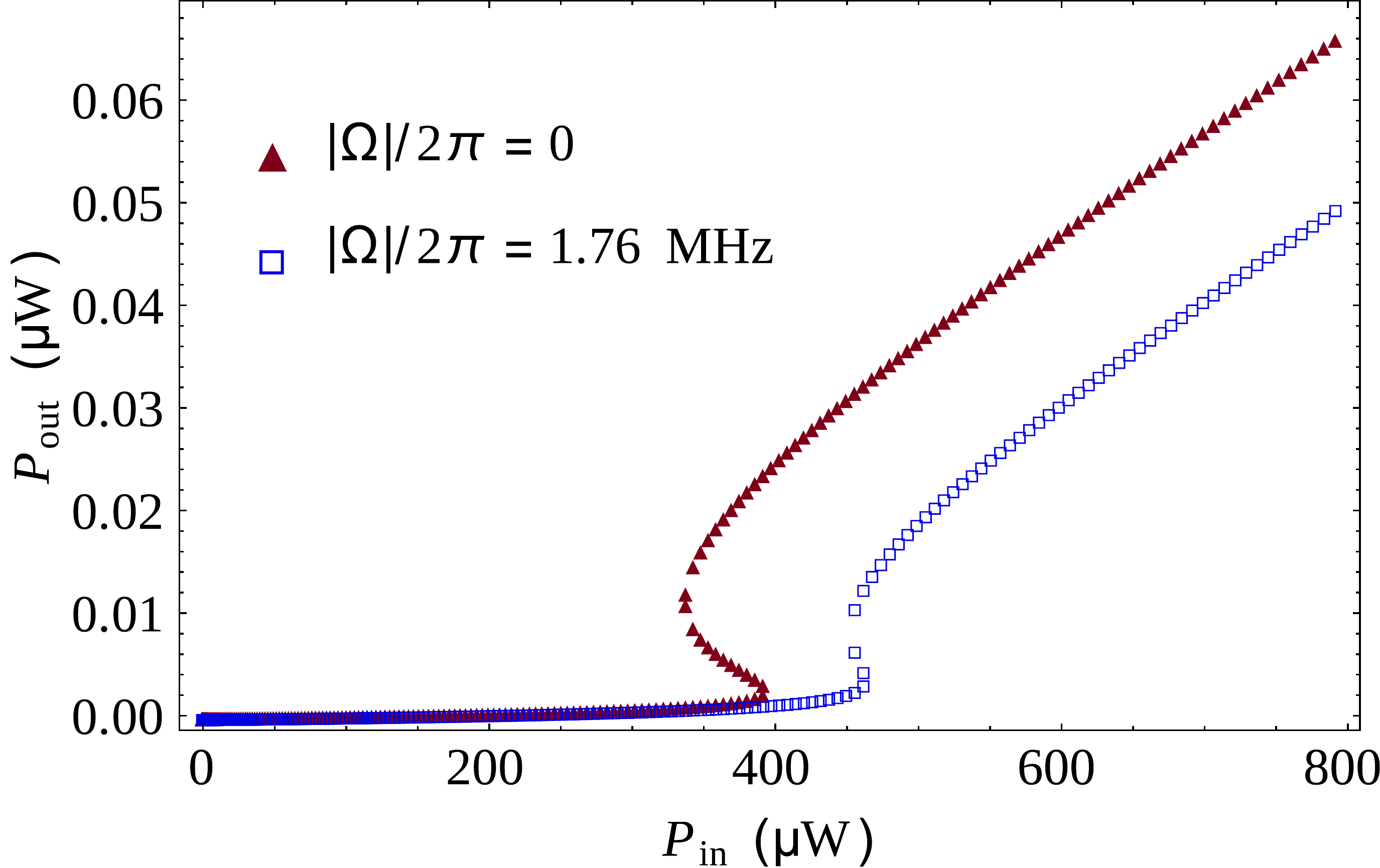}
\caption{Bistability when atoms are flowing in and out of cavity. The two curves correspond to the condition when the control light is off and on as seen in the legend.}
\label{4levelflow}
\end{figure}
At input power of 430 $\mu$W, in the case when $|\Omega| = 0 $, i.e. when there is no intersecting transverse beam, the atom-cavity system is on the upper branch of the stability curve and when $|\Omega| = 1.76$ MHz  the stability point exists only on the lower branch of the stability curve, as seen in Fig.~\ref{4levelflow}. On the lower intensity branch the output is almost zero. Hence the above fact can be used for switching on and off the cavity output by turning off and on the control beam instantaneously. For figure Fig.~\ref{4levelflow} parameters different from those used in the computation of Fig.~\ref{4bistable} are $\kappa_t = 39$ MHz, $g/2\pi = 11.8	$ kHz, $\tau_c = 0.533$ ns, $N_{at} = 5\times 10^7$, $\Delta_p= 10\Gamma$. The fraction of atoms addressed by control beam is 0.025 and $R_f/(2 \pi)= 5.5$ kHz. Above parameters are close to the experiment of Sharma et. al.~\cite{arijit}. 

\begin{figure}[b]
\centering
\includegraphics[width=7.5cm]{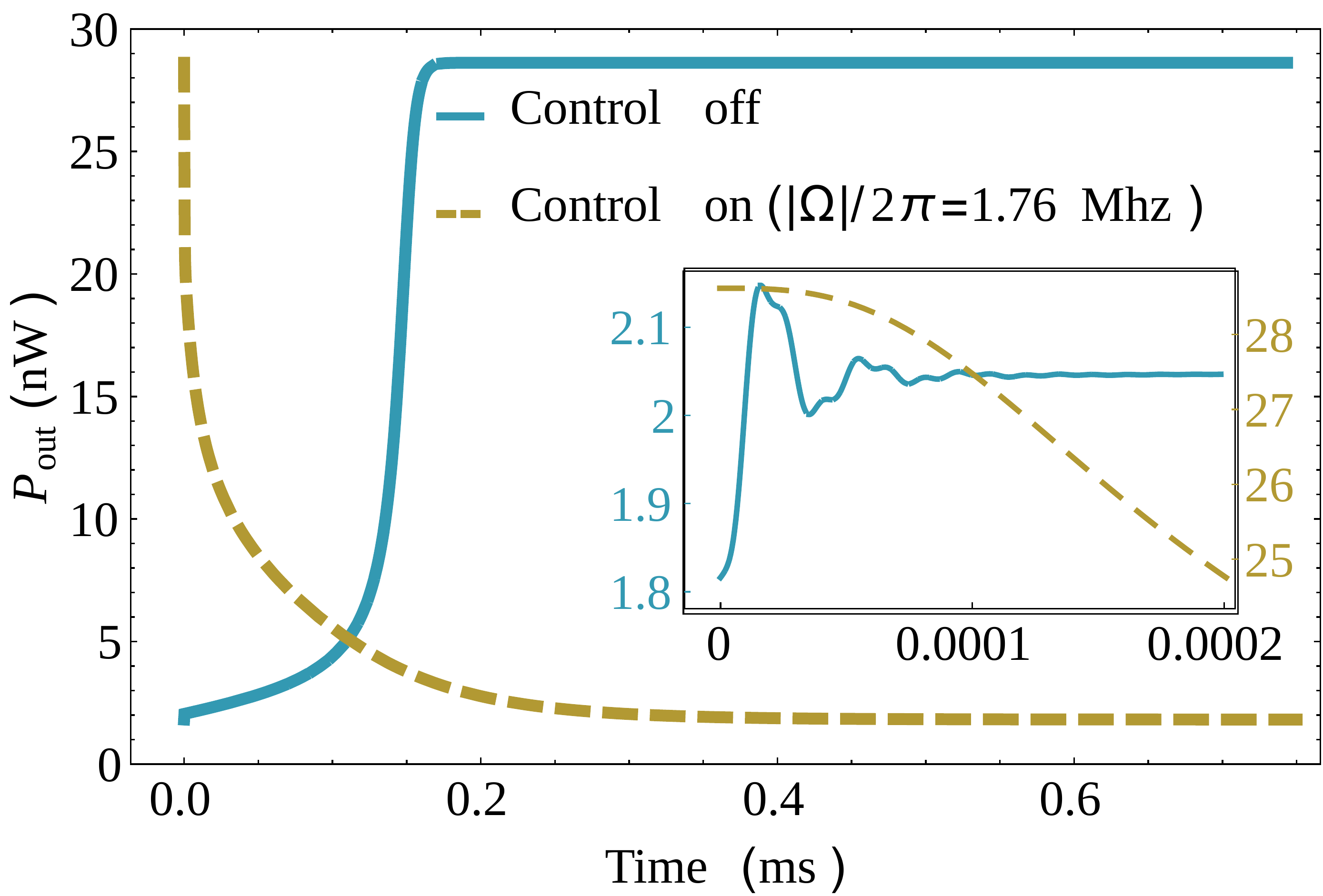}
\caption{Time evolution of cavity output power when the cavity input laser is switched by the control beam. Here control off  is $|\Omega|/2\pi = 0$  and control beam on is $|\Omega|/2\pi = 1.76 $ Mhz. The input power of 430 $\mu$W corresponds to the bistability  parameter region in Fig.~\ref{4levelflow}. The sub-ms slow response of the output power is clearly seen. The inset shows the sub-$\mu$s response of the transmitted power, where the left ordinate axis is the power in the control off case and right axis is the power for the control on. The time axis is in ms. Clearly the short time behaviour is complex.}
\label{falltime}
\end{figure}

The transient behaviour of the intra-cavity intensity, when the control beam is switched on is solved by tracking the time evolution of the atomic states in the intersection volume of the cavity mode and the transverse beam. This is done by first solving the rate equations numerically (using NDSolve in Mathematica v9.0) till the system reaches steady state with $\Omega = 0$. The second step is to solve the equations again with $\Omega \ne 0$ with initial conditions same as the final conditions of first step. Finally, the third step is to solve the equations again when the control beam is switched off ($\Omega = 0$) with initial conditions same as final state of second step. The second and third steps corresponds to switching on and off the control beam. The numerical solutions give time evolution of cavity field. The underlying physics is that the system operating point shifts from the hysteresis solution for the initial condition to that for the final state.  In this process, for parameters which closely relate to the switching regime in Sharma et. al.~\cite{arijit}, we find that the intra-cavity intensity shifts from the upper branch of the initial system when $\Omega = 0$ to the lower branch of the final system when $\Omega = 1.76$ Mhz. This is the experimentally demonstrated negative logic switching~\cite{arijit}.

For the case when the four level atoms are static, the transient response time for the switching is rapid (sub-micro-second). However, introduction of the exchange to mimic the atoms flow into and out of the cavity mode keeps readjusting the ground state populations and prevents the intra-cavity atoms from rapid optical pumping, resulting in large decay times of the cavity mode as can be seen in Fig.~\ref{falltime}. The figure shows the time evolution of output power when control beam is suddenly turned on and off. While there are fast fluctuations in the field values, the envelope of the intra-cavity intensity rise and fall times is of the order of 0.2 ms which is of the order $2\pi/R_f$. This behaviour is consistent with the long time responses measured in Sharma et. al \cite{arijit}, and the complex transient behaviour prevents the switching times in \cite{arijit} from exhibiting single exponential response.

\subsection{Two level atoms with decay loss}  \label{2levelopensec}

The special case of the four level system, where both the probe and control laser are resonant with the $\ket{1}\leftrightarrow\ket{3}$ transition, constitutes the two level system that is experimentally relevant. For such a system, there is a possibility that an atom in $\ket{3}$ decays to the other ground state $\ket{2}$, but in $\ket{2}$, the atom is no longer optically active and the overall effect of the presence of the other ground state is that of an optical loss mechanism created by the decay from $\ket{3}$. The experimental realization of this is simple and it results in a different response of the cavity transmission.

Here assuming $\Delta_c=0$ and $\gamma'_{12}=0$ and all other terms as before, the coupled differential equations describing the system can be written as,
\begin{widetext}
\begin{eqnarray}
\dot{\alpha} &=& \sqrt{\frac{\kappa_1}{\tau_c}} \alpha^{in}_p - \kappa_t \alpha  - i g N_{at} \rho_{13}\nonumber  \\
\dot{\rho}_{13} &=& -\left(\gamma_{13} + i \Delta_p \right)\rho_{13} + i g \alpha (\rho_{33} - \rho_{11})  + i \Omega (\rho_{33} - \rho_{11})+R_f(\rho^{0}_{13}-\rho_{13})\nonumber   \\
\dot{\rho}_{33} &=&  -\Gamma \rho_{33} + ig (\alpha^{*} \rho_{13} - \alpha \rho_{13}^{*} )+ i (\Omega^{*} \rho_{13} - \Omega \rho_{13}^{*} )+R_f(\rho^{0}_{33}-\rho_{33})\nonumber  \\
\dot{\rho}_{11} &=&   \frac{\Gamma}{2} \ \rho_{33} - ig (\alpha^{*} \rho_{13} - \alpha \rho_{13}^{*} ) - i (\Omega^{*} \rho_{13} - \Omega\rho_{13}^{*} ) +R_f(\rho^{0}_{11}-\rho_{11})\nonumber \\
\dot{\rho}_{22} &=&  \frac{\Gamma}{2} \ \rho_{33} +R_f(\rho^{0}_{22}-\rho_{22}) .
\end{eqnarray}
\end{widetext}
Solving for $\kappa_{at}$ in Eqn.~(\ref{alpha4level}) gives,

\begin{eqnarray}
\kappa_{at}=\frac{N_{at}gR_f A (g+\Omega/\alpha ) (B -2 i \Delta_p)}{2 B C |g  \alpha +\Omega|^2+R_f A \left[4 \Delta_p^2+B^2\right]}.
\end{eqnarray}
A,B and C are defined earlier. 

\begin{figure}[t]
\centering
\includegraphics[width=7.5cm]{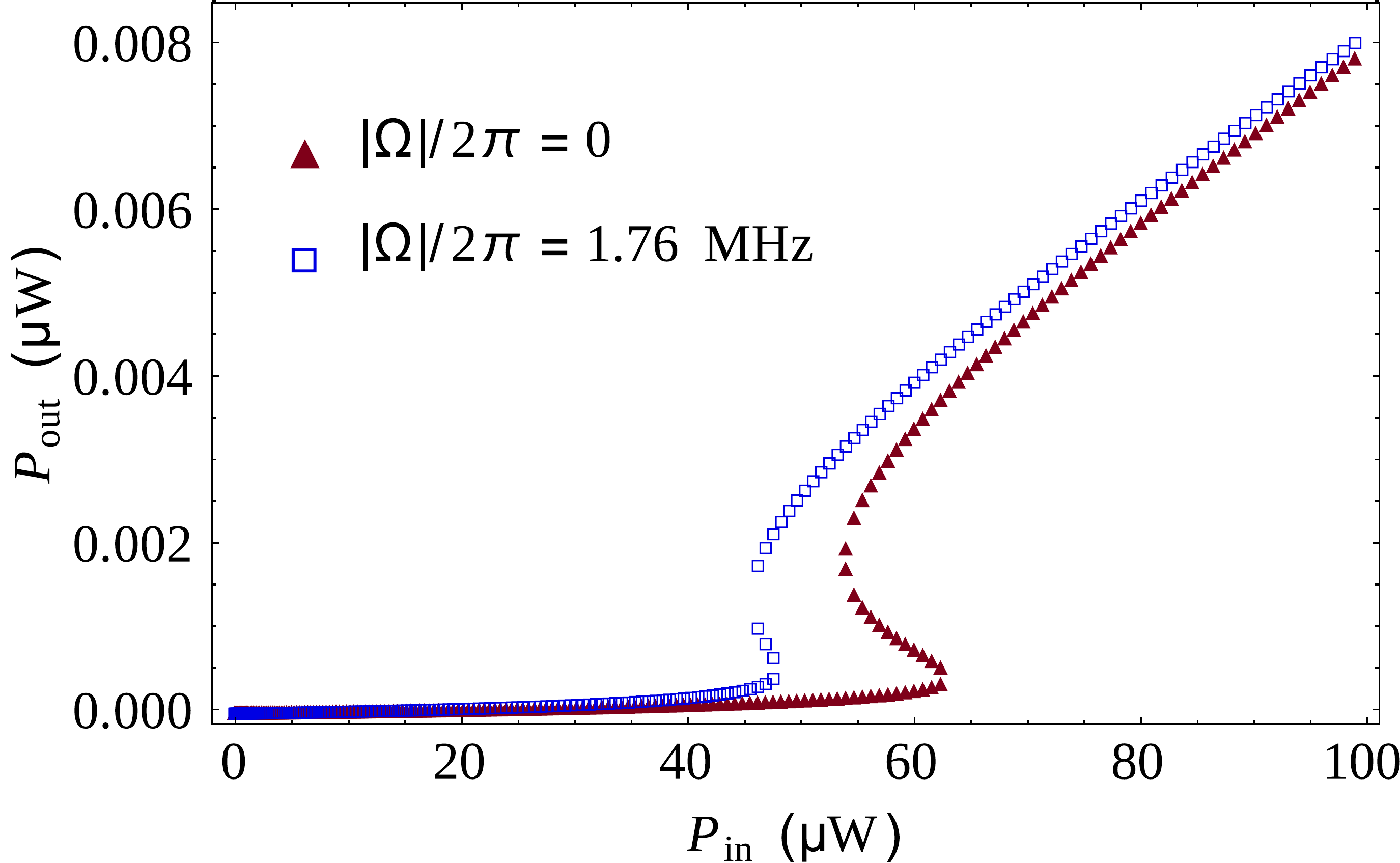}
\caption{Bistability for atoms with exchange when control beam is in resonance with levels $\ket{1}$ and $\ket{3}$, making it a two level system with loss. Here the fraction of atoms addressed by control beam is 0.2, all other parameters are same as the values used for Fig.~\ref{4levelflow}}
\label{flowreverse}
\end{figure}

Once again there exists a cubic equation in $\alpha$, which is the type of solution that supports bistability and hysteresis. Here the equations have been written with the exchange rate of atoms incorporated. Solving for the same atom cavity parameters as used for the four-level open system, we see that in this case, the control beam pumps atoms out of the cavity light cycle and hence reduce probe loss. The shift in the bistable region of the output response to lower input intensity, shown in Fig.~\ref{flowreverse} confirms this. The corresponding hysteresis plot can be readily imagined and the positive logic or cavity mode enhancement experiments~\cite{arijit} follow from the processes described above. Without undue repetition, we can observe from Fig.~\ref{flowreverse} that, in a region of input power values of the probe, the application of the appropriate intensity of control laser can promote the transmission from low intensity to high intensity in positive switching logic. The transient properties are similar that of the four-level case with non-static atoms, as can and should be expected, in line with the experimental observations.

\section{Discussion}

As can be seen from the results above, the division of the problem into a closed and open atomic ensemble interacting with the cavity mode expresses itself in the transient response to controlled intra-cavity light intensity. The theoretical solutions obtained can be explicitly tested in experiments since the values for the atomic transitions and fields used in the explicit solutions, are close to the experimental values for Rb. We expect that the generic features of the results here are independent of the particular atomic system and cavity parameters. We now discuss the simplifications and extensions of this work to experiments.
The $^{87}$Rb energy levels considered in the above calculations are the $5 S_{1/2}$, $F=1,2$ and the $5 P_{3/2}$, $F^{\prime}=1,2$, so that there are no closed optical transitions in the system. This choice represents the most general four-level system. In the experiments of Sharma et. al.~\cite{arijit}, the majority of the experiments were performed by tuning the lasers to the maxima of the Doppler broadened absorption, and therefore close to the $5 P_{3/2}$, $F^{\prime}=3,4$. Further the Doppler spread and the velocity dependent coupling of the atoms to the cavity in the open system case has not been considered in the model, which is expected to contribute to the details of the experimental results. While the effort has been to keep the agreement between theoretical parameters and experiment as close as possible, there are significant departures between the analysis in this article and the experimental realization  in Sharma et. al \cite{arijit}. Therefore while qualitative agreement can be expected between the theory here and the experiment~\cite{arijit}, the present work does not attempt to provide detailed quantitative agreement.

\subsection{Closed system of atoms}

\begin{figure}[t]
\centering
\includegraphics[width=7.5cm]{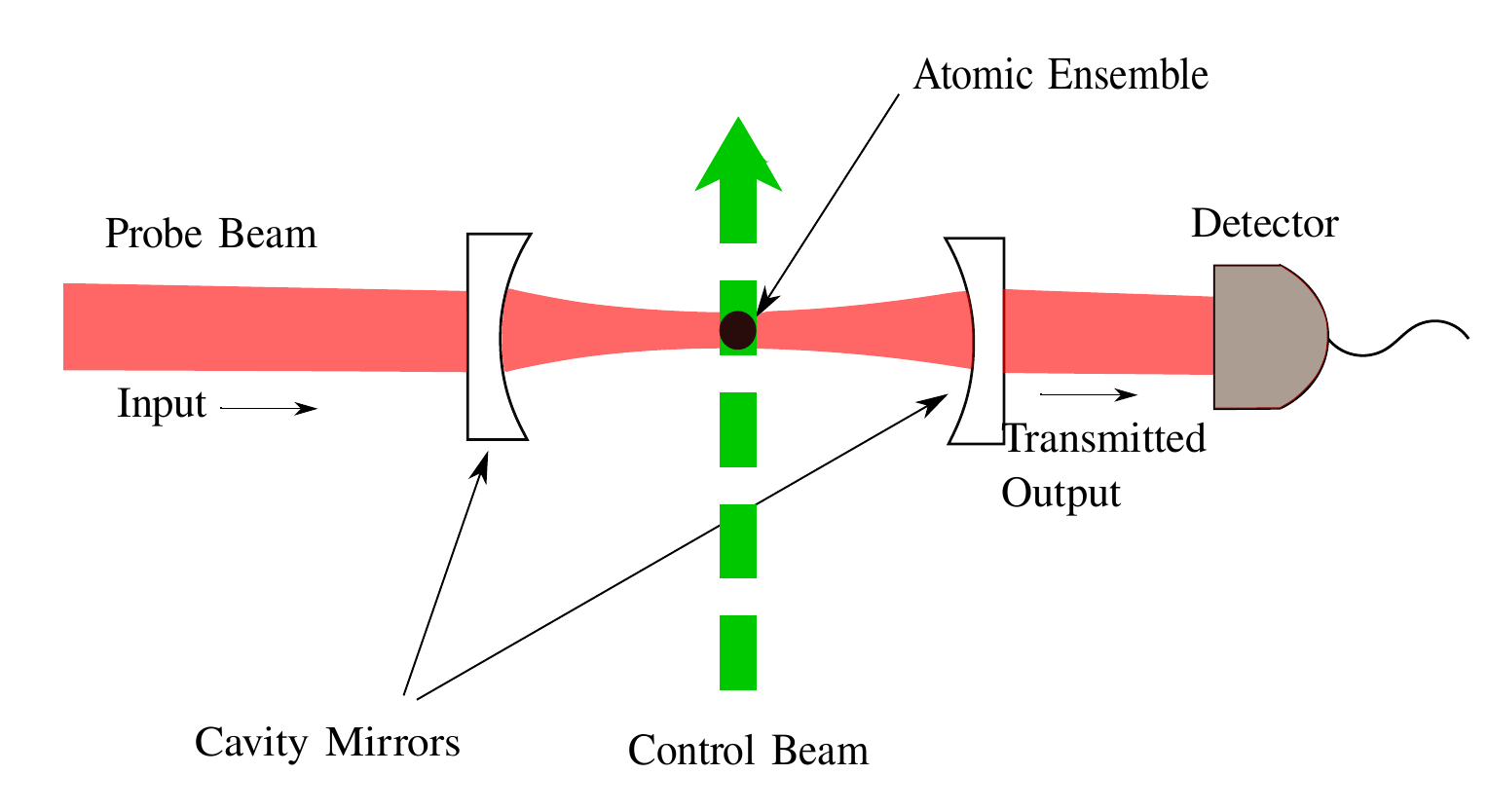}
\caption{Schematic diagram for closed system of atoms. The black spot can represent an ensemble of laser cooled and trapped atoms or ions, or atoms in a dipole trap.}
\label{closed_experiment}
\end{figure}

A viable candidate for a closed system of atoms can be a cold atom ensemble that is contained within the mode volume of the cavity as illustrated in the Fig.~\ref{closed_experiment}. For such a system, using initially state prepared atoms, a number of situations with four, three and two level atomic systems can be studied, with a small number of atoms present in the cavity mode. The simulations for this case above have been done with $N_{at}= 5\times10^4$ atoms and the lasers plus the detection technologies required are readily available. In all cases with the closed system, resonant light in the cavity mode will lead to optical pumping into the alternate ground state and lead the system into transparency. Therefore to control the intra-cavity light field, it is essential to have a non-zero transverse field which allows the atom to interact with the cavity field. 

When a weak control light is incident on the atoms connecting an excited state with the erstwhile dark ground state, the optical pumping effects are reversed and a simultaneous non-zero population of both ground states manifests. In a specific range of system parameters which depend on the detunings and intensities of the two fields, steady state bistability of the transmitted light intensity is observed. This can be used to control the transmitted light intensity through the cavity and in the limit of weak fields and small atom numbers, can be used as a very sensitive all optical switch. The choice of doing this with either the four level atom or the three level atom exists.

\begin{figure}[t]
\centering
\includegraphics[width=7.5cm]{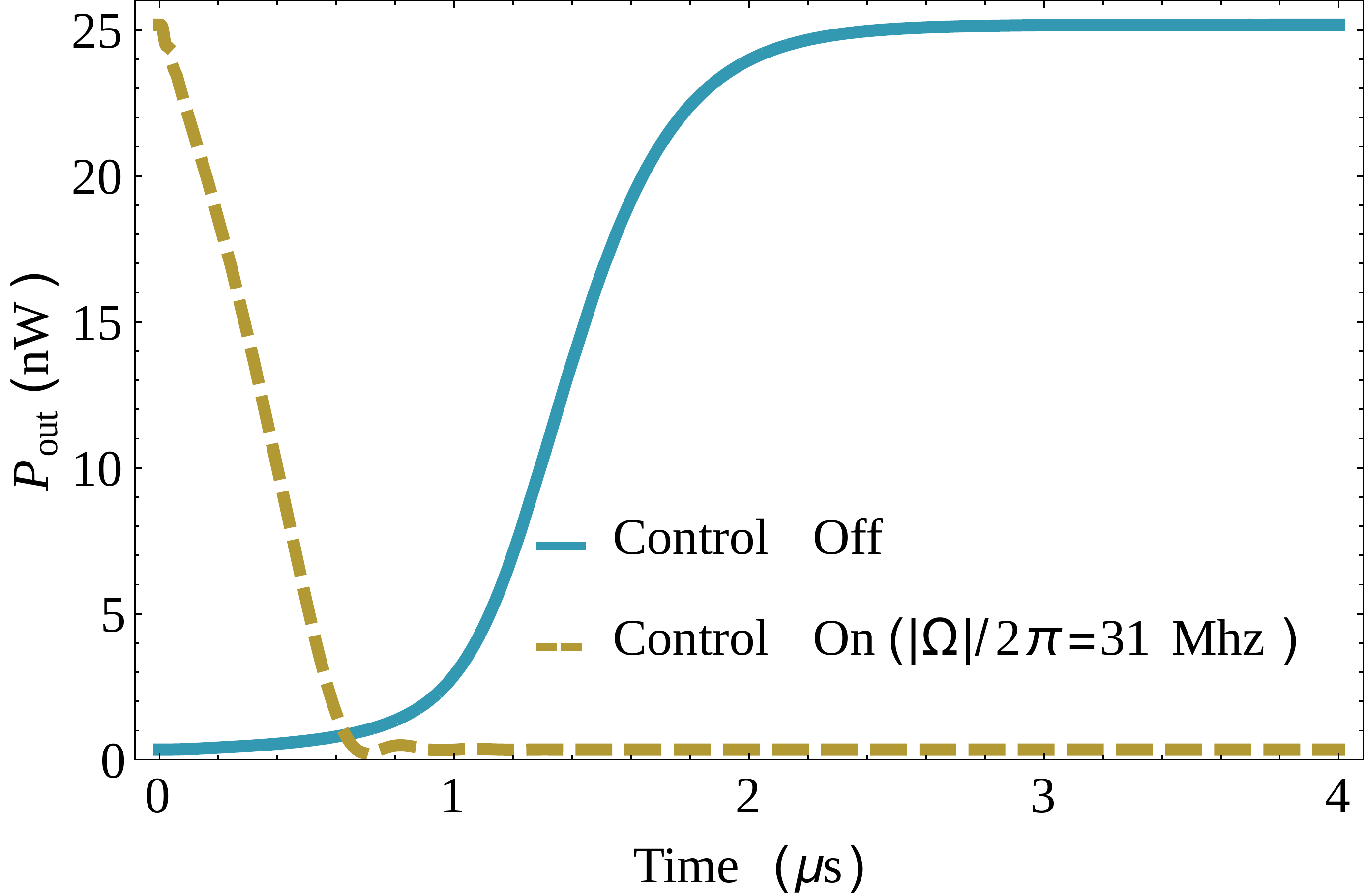}
\caption{Fast Switching times for closed 4 level system. Here control off  is $|\Omega|/2\pi = 0$  and control beam on is $|\Omega|/2\pi = 31 $ Mhz. The input power of 1 $\mu$W corresponds to the lower part of the second bistability curve in Fig.~\ref{4shift} when control is on and the no atoms curve in Fig.~\ref{4bistable} when control is off.}
\label{fasttimes}
\end{figure}

Due to negligible Doppler broadening for laser cooled atoms, the three level system is practically realizable. The solution for this system is particularly rich, as seen in~\ref{3levelsec} . As the relative detunings are altered, complex output to input intensity dependences show up, resulting in multiple bistability in some cases. The parameter space for observation of multiple bistability needs careful adjustment of intensity and detunings, an example of this is seen in Fig.~\ref{3levelbi} and Fig.~\ref{3levelhys}. It might be possible to exploit the multiple bistability regime for a 2$^+$-level all optical switch, where instead of just the turning on and off of the cavity field, intermediate, stable intra-cavity light intensities are possible. This would open up new possibilities for the precise control of the degree of atom-cavity field interaction, by manipulation of the control light field.

The transient response of the intra-cavity light field for the closed system depends on the atom-cavity coupling, the reflectivity of the cavity mirrors and the cavity mode volume. The transient response in this case is of the order of a microsecond (Fig~\ref{fasttimes}), for both the rise and fall times of the intra-cavity intensity, when solved for the experimental parameters of Ray et. al.~\cite{tridib}. A significant fraction of the physics discussed here is possible on our experiment with cold atoms in the cavity~\cite{tridib}. While a cold atom ensemble also has its losses and flows, our ability to state prepare the system allows it to be used partially in the manner treated in the present manuscript.

\subsection{Open system of atoms}

When the atoms from the reservoir move in and out of the cavity mode volume, the atomic subsystem is open and how this exchange of atoms affects the experimental measurement can be studied in this vapour cell in the cavity (Fig.~\ref{twobeams}) setup. The cell can be placed in either a ring cavity \cite{WangPRA65,joshi}, or a standing wave Fabry-P\'erot cavity \cite{arijit,zou-zhu,MlynekPRA}, as discussed here.

In this system we have a significantly larger number of atoms in the cavity mode at any given time, consistent with the vapour pressure at room temperature. In addition the cavity decay rates for photons are much higher since the cavity has very high loss, i.e. $\kappa_t$ is very high. Further the atom-cavity coupling $g$ is much lower in this case than that for the cold atoms.
\begin{figure}[t]
\centering
\includegraphics[width=7.5cm]{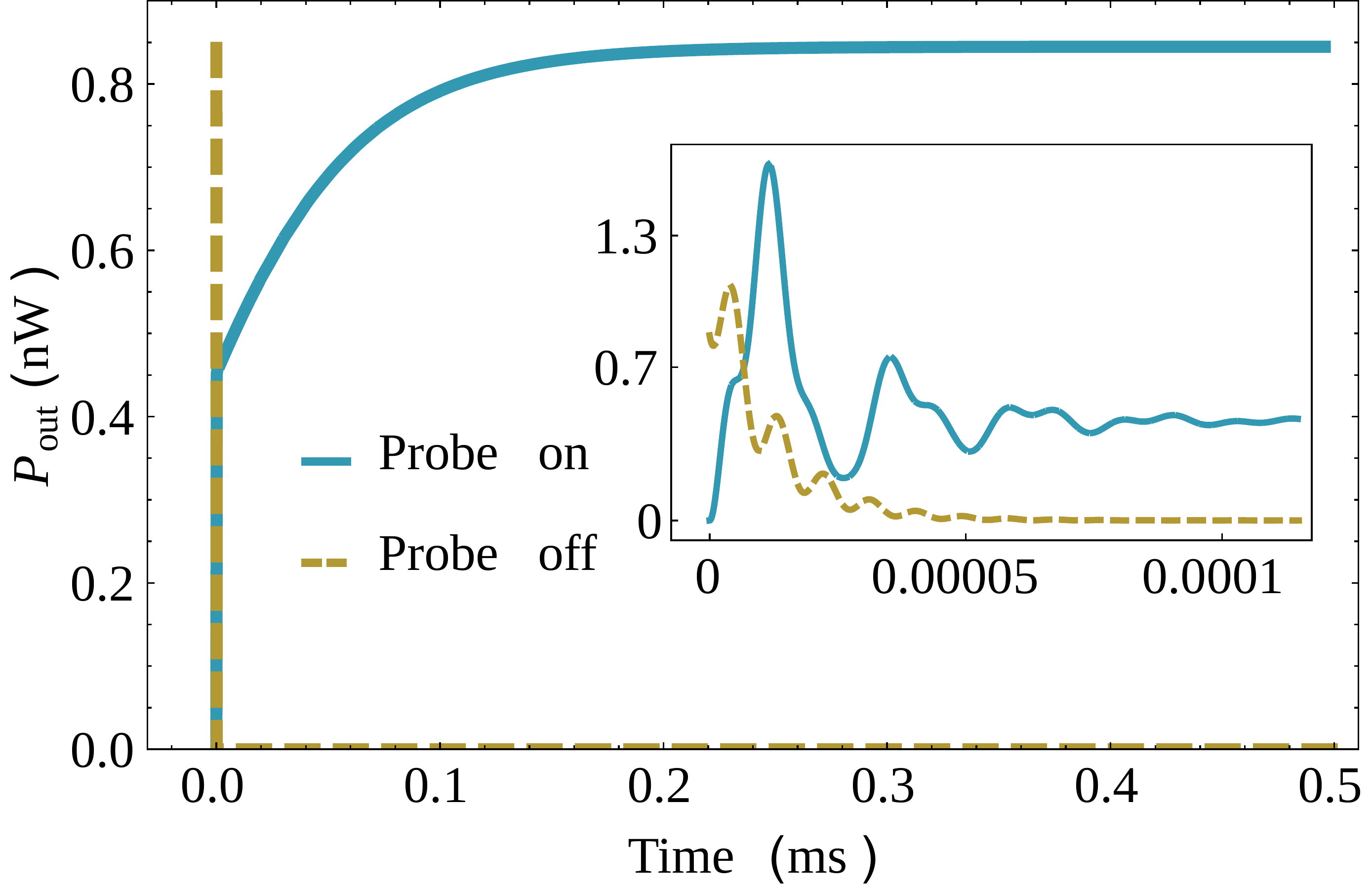}
\caption{Rise and decay of transmitted intensity when cavity laser is switched on and off respectively. The input power of 300 $\mu$W corresponds to the lower branch of hysteresis curve in Fig.~\ref{4levelflow} with $\Omega = 0$. }
\label{riselower}
\end{figure}

Of the many connections that the model makes with the experiment~\cite{arijit}, a key one is in the understanding of the long rise time ($\approx 0.2$ ms) for the cavity field, when resonant probe light is suddenly switched on, in the absence of control beam. This long rise time is due to the inclusion of atom flow in and out from the cavity mode volume, as seen in Fig.~\ref{riselower}. If the flow of atoms is not incorporated, then the field build up in the cavity is two orders of magnitude faster (2 $\mu$s). When the probe field is physically switched off, the decay of the cavity field is very rapid. Both the rise and the decay of the resonant probe field is shown in Fig.~\ref{riselower}. In both the rise and decay, fast oscillations are seen in the time evolution. From this it is clear that the transient response of the resonant atom-cavity system is critically altered by the movement of atoms in and out of the cavity mode. The constant $R_f$ for the calculations is extracted from the experiment~\cite{arijit} and then introduced into the evolution equations for the atom-field system to obtain the results reported here.

For the open system, even when the control beam is off, the atom-cavity bistability is observed~\cite{arijit}. This is due to the continuous flow of ground state atoms in the cavity mode volume, which has the effect of continuously maintaining the atomic population in both atomic ground states. The effect of the control beam is that the particular value of input light intensity, where the cavity field solution becomes bistable shifts, (a) to higher input light power values, when the control is on the complementary transition (full four level system), and (b) to lower light input powers when the control is on the same transition as the probe (two level system with decay).
 
The observation that for high intra-cavity intensity of the probe light, even relatively large intensities of control light is unable to switch the probe transmission off, is explained by the intracavity intensity jumping between the upper branches of the hysteresis curve, resulting in small changes in the intra-cavity intensity, and therefore the transmitted light intensity. It is therefore clear that the switching of the cavity light intensity as described in Sharma et. al.~\cite{arijit} is possible only in a window of parameter space, which can be calculated using the methods described here.	

A particularly challenging problem posed by the experiment~\cite{arijit} was the large time constant for the decay of the cavity mode intensity, when the control beam is switched on, in the negative logic case. The measured time constant was $\approx 0.5$ ms. This was explained by the incorporation of the exchange of atoms between the cavity mode and the thermal reservoir. The decay time for the cavity field on transverse switching, for the open system is shown in Fig.~\ref{falltime}. This large decay time reduces by orders of magnitude(0.6 $\mu$s) for the closed system.

\section{Conclusions}

The atom-cavity system's ability to transmit resonant light is studied in detail. The solutions for 4-level, 3-level and 2-level atoms are solved for realistic atom-cavity parameters. The solutions are constructed for the case when the atoms are stationary in the cavity mode and in generalization, when the atoms are exchanged with a reservoir. It is shown that the motion of the atoms is reflected in the transient properties of the transmitted light through the cavity. All the key features of the recent experiment of Sharma et. al.~\cite{arijit} are qualitatively understood on the basis of the theoretical analysis here. The possibility of adapting these systems for the study of multiple-bistability, multistability and with cold atom ensembles are exciting prospects for the future experiments.   

\section*{Acknowledgement}
We thank Jose Luis Gomez-Mu\~noz for the QUANTUM add-on for Mathematica, which was used here. Tridib Ray and Andal Narayanan are acknowledged for discussions.

\begin{widetext}

\appendix*
\section{}
\subsubsection{3 level atoms}

The values of $T_1, ..., T_6$ in~\ref{3levelsec},

\begin{eqnarray}
T_1 &=& 2g^2N_{at}|\Omega|^2\left[\Gamma  (\Gamma -2 i \Delta_p) \left({\gamma'}_{12}^2+(\Delta_p-\Delta_c)^2\right)+2 \Gamma  |\Omega|^2 (\gamma'_{12}+i\Delta_p-i\Delta_c)\right]\nonumber \\
T_2 &=& 4g^4N_{at}|\Omega|^2\left[-2\gamma'_{12} (i\Delta_p+i\Delta_c)+\Gamma  (\gamma'_{12}+i\Delta_p-i\Delta_c)\right]\nonumber \\
T_3 &=& 4 \Gamma g^6\nonumber \\
T_4 &=&  g^4 \left[\Gamma^2 \gamma'_{12}+2\Gamma \Delta_c (\Delta_p-\Delta_c)+3 |\Omega|^2 (\Gamma +4\gamma'_{12})\right]\nonumber \\
T_5&=&g^2 \left[\Gamma  \left(\Gamma ^2+4 \Delta_c^2\right) \left({\gamma'}_{12}^2+(\Delta_p-\Delta_c)^2\right)+8 |\Omega|^2 \left(\Gamma ^2 \gamma'_{12}+ 3\Gamma  {\gamma'}_{12}^2+2\Gamma  (\Delta_p-\Delta_c)^2+\gamma'_{12} (\Delta_p+\Delta_c)^2\right)\right.  \nonumber \\
&&+ \left. 12 |\Omega|^4  (\Gamma +4 \gamma'_{12})\right]\nonumber \\
T_6 &=&\Gamma  |\Omega|^2 \left(\Gamma ^2+4 \Delta_p^2\right) \left({\gamma'}_{12}^2+(\Delta_p-\Delta_c)^2\right)+4 \Gamma  |\Omega|^6+|\Omega|^4[ \Gamma^2  \gamma'_{12}+2 \Gamma \Delta_p (\Delta_c- \Delta_p)].
\end{eqnarray} 


\subsubsection{Open system of four level atoms}
The coupled time dependent equations including the flow for the case in~\ref{4levelopensec},

 \begin{eqnarray}
\dot{\alpha} &=& \sqrt{\frac{\kappa_1}{\tau_c}} \alpha^{in}_p - \kappa_t \alpha  - i g N_{at} \rho_{13}  \nonumber \\
\dot{\rho}_{13} &=& -\left(\gamma_{13} + i \Delta_p \right)\rho_{13} + i g \alpha (\rho_{33} - \rho_{11}) +R_f(\rho^{0}_{13}-\rho_{13})   \nonumber \\
\dot{\rho}_{33} &=&  -\Gamma \rho_{33} + ig (\alpha^{*} \rho_{13} - \alpha \rho_{13}^{*} )+R_f(\rho^{0}_{33}-\rho_{33}) \nonumber  \\
\dot{\rho}_{11} &=&  \frac{\Gamma}{2} \ \rho_{44} + \frac{\Gamma}{2} \ \rho_{33} - ig (\alpha^{*} \rho_{13} - \alpha \rho_{13}^{*} ) +R_f(\rho^{0}_{11}-\rho_{11}) \nonumber \\
\dot{\rho}_{24} &=& -\left(\gamma_{24} + i \Delta_c \right)\rho_{24} + i \Omega (\rho_{44} - \rho_{22}) +R_f(\rho^{0}_{24}-\rho_{24})  \nonumber \\
\dot{\rho}_{44} &=&  -\Gamma \rho_{44} + i  (\Omega^{*} \rho_{24} - \Omega \rho_{24}^{*} )+R_f(\rho^{0}_{44}-\rho_{44}) \nonumber \\
\dot{\rho}_{22} &=&  \frac{\Gamma}{2} \ \rho_{44} + \frac{\Gamma}{2} \ \rho_{33} - ig (\Omega^{*} \rho_{24} - \Omega \rho_{24}^{*} )  +R_f(\rho^{0}_{22}-\rho_{22}).
\end{eqnarray}

For $\Omega \ne 0$ we get $\kappa_{at} = \frac{E}{F}$ where,

\begin{eqnarray}
E&=&g^2N_{at} A (B -2 i \Delta_p) \left\{4 |\Omega|^2 B^2+R_f A\left[4 \Delta_c^2+B^2\right]\right\} \nonumber\\
F&=& 2 |\alpha|^2g^2 B \left\{16 |\Omega|^2 B^2+A C \left[4 \Delta_c^2+B^2\right]\right\}+A \left(4 \Delta_p^2+B^2\right) \left\{2 |\Omega|^2 B C+R_f A \left[4 \Delta_c^2+B^2\right])\right\} 
\end{eqnarray}

$A, B$ and $C$ are defined in~\ref{4levelopensec}.


\subsubsection{Open system of two level atoms with decay loss}

The equations for the case in~\ref{2levelopensec},

\begin{eqnarray}
\dot{\alpha} &=& \sqrt{\frac{\kappa_1}{\tau_c}} \alpha^{in}_p - \kappa_t \alpha  - i g N_{at} \rho_{13} \nonumber \\
\dot{\rho}_{13} &=& -\left(\gamma_{13} + i \Delta_p \right)\rho_{13} + i g \alpha (\rho_{33} - \rho_{11})  + i \Omega (\rho_{33} - \rho_{11})+R_f(\rho^{0}_{13}-\rho_{13}) \nonumber  \\
\dot{\rho}_{33} &=&  -\Gamma \rho_{33} + ig (\alpha^{*} \rho_{13} - \alpha \rho_{13}^{*} )+ i (\Omega^{*} \rho_{13} - \Omega \rho_{13}^{*} )+R_f(\rho^{0}_{33}-\rho_{33}) \nonumber  \\
\dot{\rho}_{11} &=&   \frac{\Gamma}{2} \ \rho_{33} - ig (\alpha^{*} \rho_{13} - \alpha \rho_{13}^{*} ) - i (\Omega^{*} \rho_{13} - \Omega\rho_{13}^{*} ) +R_f(\rho^{0}_{11}-\rho_{11}) \nonumber \\
\dot{\rho}_{22} &=&  \frac{\Gamma}{2} \ \rho_{33} +R_f(\rho^{0}_{22}-\rho_{22}) .
\end{eqnarray}

\end{widetext}



\begin{thebibliography}{}

\bibitem{arijit}
Arijit Sharma, Tridib Ray, Rahul V. Sawant, G. Sheikholeslami, S. A. Rangwala, and D. Budker, \textit{Phys. Rev. A} \textbf{91}, 043824 (2015).

\bibitem{gibbs}
H. M. Gibbs. \textit{Optical Bistability: Controlling Light with Light} (Academic, New York, 1985).

\bibitem{revLugiatoGibbs}
L. A. Lugiato, in Progress in Optics, edited by E. Wolf (North Holland, Amsterdam, 1984), XXI, 70-204, 1984.

\bibitem{WangPRA65}
H. Wang, D. J. Goorskey, and M. Xiao, \textit{Phys. Rev. A} \textbf{65}, 011801(R) (2001).

\bibitem{WangPRA65051802}
H. Wang, D. Goorskey, and M. Xiao, \textit{Phys. Rev. A} \textbf{65}, 051802(R) (2002).

\bibitem{MlynekPRA}
J. Mlynek, F. Mitschke, R. Deserno, and W. Lange, \textit{Phys. Rev. A} \textbf{29}, 1297 (1984)

\bibitem{LowenauPRL}
J. P. Lowenau, S. Schmitt-Rink, and H. Haug, \textit{Phys. Rev. Lett.} \textbf{49}, 1511 (1982).

\bibitem{OrozcoPRA}
L. A. Orozco, H. J. Kimble, A. T. Rosenberger, L. A. Lugiato, M. L. Asquini, M. Brambilla, and L. M. Narducci, \textit{Phys. Rev. A} \textbf{39}, 1235 (1989).

\bibitem{macovei}
Viorel Ciornea and Mihai A. Macovei, \textit{Phys. Rev. A} \textbf{90}, 043837 (2014).

\bibitem{raimond}
J. M. Raimond, M. Brune, and S. Haroche, \textit{Rev. Mod. Phys.} \textbf{73}, 565 (2001).

\bibitem{Haroche}
Serge Haroche, Jean-Michel Raimond, \textit{Exploring the Quantum: Atoms, Cavities, and Photons} (Oxford University Press, Oxford, 2013)

\bibitem{rempe}
Tatjana Wilk, Simon C. Webster, Axel Kuhn, and Gerhard Rempe, \textit{Science}, \textbf{317}, 5837 (2007)

\bibitem{walther}
Matthias Keller, Birgit Lange, Kazuhiro Hayasaka, Wolfgang Lange1, and Herbert Walther, \textit{Nature}, \textbf{431}, 1075-1078 (2004)

\bibitem{kimble}
J. McKeever, A. Boca, A. D. Boozer, R. Miller, J. R. Buck, A. Kuzmich, and H. J. Kimble, \textit{Science}, \textbf{303}, 5666 (2004) 
\bibitem{Cha03}
H. W. Chan, A. T. Black and V. Vuleti$\acute{c}$, \textit{Phys. Rev. Lett.} \textbf{90}, 063003 (2003).

\bibitem{Her07}
G. Hernandez, J. Zhang and Y. Zhu, \textit{Phys. Rev. A} \textbf{76}, 053814 (2007).

\bibitem{Bie10}
T. Bienaime, S. Bux, E. Lucioni, Ph. W. Courteille, N. Piovella and R. Kaiser, \textit{Phys. Rev. Lett.}, \textbf{104}, 183602 (2010).\par

\bibitem{Wan12}
Y. Wang, J. Zhang and Y. Zhu, \textit{Phys. Rev. A} \textbf{85}, 013814 (2012).

\bibitem{tridib}
Tridib Ray, Arijit Sharma, S. Jyothi, and S. A. Rangwala, \textit{Phys. Rev. A} \textbf{87}, 033832 (2013).

\bibitem{ming}
L\"u Xin-You, Li Jia-Hua, Liu Ji-Bing and Luo Jin-Ming, \textit{J. Phys. B: At. Mol. Opt. Phys.}, \textbf{39}, 5161-5171 (2006) 

\bibitem{anton}
M.A. Ant\'on, Oscar G. Calder\'on , Sonia Melle, I. Gonzalo , F. Carre\~no, \textit{Optics Communications}, \textbf{268}, 146-154 (2006) . 

\bibitem{scully}
Marlan O. Scully and M. Suhail Zubairy, \textit{Quantum Optics} (Cambridge University Press, 1997)

\bibitem{walls}
D. F.Walls \& G. J.Milburn, \textit{Quantum Optics} (Springer-Verlag, Berlin, Heidelberg, 1994).

\bibitem{albert}
Magnus Albert, Ph.D. thesis, Danish National Research Foundation Center for Quantum Optics-QUANTOP, Department of Physics and Astronomy, The University of Aarhus (2010).

\bibitem{albert1}
M. Albert, J. P. Marler, P. F. Herskind, A. Dantan, and M. Drewsen, \textit{Phys. Rev. A} \textbf{85}, 023818 (2012)


\bibitem{steck}
Daniel A. Steck, “Rubidium 87 D Line Data,” available online at http://steck.us/alkalidata (revision 2.1.4, 23 December 2010).

\bibitem{boyd}
R. W. Boyd, \textit{Nonlinear Optics} (Academic, NY, 2008), 3nd ed., Chap. 7.

\bibitem{zou-zhu}
Bichen Zou and Yifu Zhu, \textit{Phys. Rev. A} \textbf{87}, 053802 (2013)

\bibitem{EITrev}
 M. Fleischhauer, A. Imamoglu, and J. P. Marangos, \textit{Rev. Mod. Phys}, \textbf{77}, 633 (2005).

\bibitem{joshi}
Amitabh Joshi, Andy Brown, Hai Wang, and Min Xiao, \textit{Phys. Rev. A} \textbf{67}, 041801(R) (2003)

\bibitem{Wei-Zhu}
Xiaogang Wei, Jiepeng Zhang, and Yifu Zhu, \textit{Phys. Rev. A} \textbf{82}, 033808 (2010)
\end{thebibliography}
\end{document}